\documentclass[mypaper,10pt,twoside]{CoAst}
\usepackage{epsf,color,graphicx,picinpar,fancyhdr,multicol,multirow,latexsym,fancybox,amssymb,supertabular,longtable,rotating,nicefrac,amsmath}
\usepackage{lscape} 
\usepackage{url} 
\usepackage{sfmath}

\usepackage{ucs}
\usepackage{booktabs}
\usepackage{inputenc}
\usepackage{fontenc}
\usepackage{dcolumn}

\usepackage{natbib}

\renewcommand{\chaptermark}[1]{\markboth{#1}{}}

\renewcommand{\headrulewidth}{0pt}
\def\setpubdate{2010}
\newcommand{\kopf}{\small\itshape 
Comm. in Asteroseismology - Complementary Topics \\ 
Volume 1, \setpubdate \\
\copyright~Austrian Academy of Sciences}

\newcommand{\Authors}[1]{\begin{center}\normalsize\bf\sf #1 \end{center}}

\renewcommand{\author}[1]{\begin{center}\normalsize\bf\sf #1 \end{center}}
\newcommand{\Address}[1]{\begin{center}\small\sf #1 \end{center}}

\renewenvironment{abstract}{\section*{Abstract}\normalsize\sf}{}
\newcommand{\References}[1]{\begin{flushleft}{\large References\\}\vspace*{2mm}\small #1 \end{flushleft}}

\newcommand{\chapterCoAst}[2]
{\chapter[\sf\normalsize #1\\ 
\footnotesize \hspace*{5mm}by #2 \sf\normalsize][]
{#1\\}
\rhead[\fancyplain{}{\sf\footnotesize \center{#1}}]{\fancyplain{}{\sffamily\thepage}}
\lhead[\fancyplain{\kopf}{\sffamily\thepage}]{\fancyplain{\kopf}{\sf\footnotesize \center{#2}}}}

\newcommand{\figureDSSN}[5]{\begin{figure}[#4]
\centering
\includegraphics*[#5]{#1}
\caption{#2}
\label{#3}
\end{figure}}

\renewcommand{\chaptername}{}

\newcommand{\acknowledgments}[1]{\vspace*{5mm}\noindent  \textbf{Acknowledgments.} #1}

\begin{document}

\pagestyle{empty}

\sf


\setcounter{page}{1}
\thispagestyle{empty}
\vspace*{-1cm}
\begin{center}


\huge\sf    Communications in Asteroseismology\\
Complementary Topics\\
\vspace*{1cm}
\large 
   Volume 1 \\ \the\year\\
\vspace*{4cm}
\vspace*{4mm}
\vspace*{5cm}

\begin{figure}[!ht]
\centering
\includegraphics*[width=60mm,clip]{OEAW_englisch_2009_SW}

\end{figure}

\normalsize
\end{center}
\newpage
\thispagestyle{empty}
\normalsize
\vspace*{-1cm}

\begin{center}
\begin{large}Communications in Asteroseismology - Complementary Topics\end{large}\\
Editor-in-Chief: \textbf{Michel Breger}, michel.breger@univie.ac.at\\
Editorial Assistant: \textbf{Isolde M\"uller}, isolde.mueller@univie.ac.at\\
Layout \& Production Manager: \textbf{Isolde M\"uller}, isolde.mueller@univie.ac.at\\

\vspace{2mm}
CoAst and CoAct Editorial and Production Office\\
T\"urkenschanzstra\ss e 17, A - 1180 Wien, Austria\\
\textit{http://www.oeaw.ac.at/CoAst/ \\ Comm.Astro@univie.ac.at\\}
\vspace*{4mm}
\textbf{Editorial Board:} Conny Aerts, Gerald Handler, \\ Don Kurtz, Jaymie Matthews, Ennio Poretti\\

\vspace*{1.5cm}

\begin{large}Cover Illustration\end{large}\\
\end{center}

\vspace*{1.5cm}
\begin{center}
\small
\sf British Library Cataloguing in Publication data.\\
\sf A Catalogue record for this book is available from the British Library.
\end{center}
\vspace*{1.4cm}
\small
\begin{center}
All rights reserved\\
ISBN ???-?-????-????-?\\
ISSN ????-????\\
Copyright~\copyright~2010 by\\
Austrian Academy of Sciences \\
Vienna\\
\vspace*{2mm}
Austrian Academy of Sciences Press\\
A-1011 Wien, Postfach 471, Postgasse 7/4\\
Tel. +43-1-515 81/DW 3402-3406, +43-1-512 9050\\ Fax +43-1-515 81/DW 3400\\
http://verlag.oeaw.ac.at, e-mail: verlag@oeaw.ac.at\\
\end{center}

		\pagestyle{fancyplain}
		\setlength{\columnseprule}{0.2pt}
		\normalsize



\newpage
\thispagestyle{empty}
\normalsize

\sf
\chapterCoAst{Combine User's Manual}{P. Reegen}

\Authors{P. Reegen$^1$} 
\Address{$^1$ Institut f\"ur Astronomie, T\"urkenschanzstrasse 17, 1180 Vienna, Austria\\
reegen@astro.univie.ac.at}

\noindent
\begin{abstract}
{\sc Combine} is an add-on to {\sc SigSpec} and {\sc Cinderella}. A {\sc SigSpec} result file or a file generated by {\sc Cinderella} contains the significant sinusoidal signal components in a time series. In this file, {\sc Combine} checks one frequency after the other for being a linear combination of previously examined frequencies. If this attempt fails, the corresponding frequency is considered ``genuine''. Only genuine frequencies are used to form linear combinations subsequently. A purely heuristic model is employed to assign a reliability to each linear combination and to justify whether to consider a frequency genuine or a linear combination.
\end{abstract}

\section{What is {\sc Combine}?}\label{COMBINE_What is}

{\sc Combine} performs an iterative analysis of the frequencies in a {\tt result.dat} file generated by {\sc SigSpec} (Reegen 2005, 2007, 2009) or one of the output files generated by {\sc Cinderella} (Reegen et al.~2008; Reegen~2009). The input file type is detected automatically.

If the attempt to interpret a given frequency as a linear combination fails, this frequency is considered genuine. Only genuine frequencies are used to form linear combinations in the subsequent iterations. The decision whether to accept a linear combination is drawn using a mathematical model to assign an equivalent spectral significance (hereafter abbreviated by `sig') to a linear combination. This equivalent sig is compared to the sig of the given signal component, and only if it is high enough, the program adopts it.

If there is more than one linear combination available, {\sc Combine} picks the one with the highest equivalent significance.

The underlying model leading to equivalent sigs and the reliabilities of linear combinations is purely heuristic and attempt to mimic the examination by an experienced person.

\section{Input}\label{COMBINE_Input}

{\sc Combine} is called by the command line

\begin{scriptsize}\begin{verbatim}
combine <infile>
\end{verbatim}\end{scriptsize}

\noindent where {\tt <infile>} is the name (or path, if desired) of a {\sc SigSpec} result file or an output file generated by {\sc Cinderella}.

\vspace{12pt}
{\bf Caution: {\sc Combine} overwrites existing output files!}
\vspace{12pt}

Furthermore, the user may pass a set of specifications to {\sc Combine} by means of a file {\tt <infile>.ini} in the same folder as {\tt <infile>}. For specifications not given by the user, defaults are used.

\vspace{12pt}
{\bf The file {\tt <infile>.ini} has to be terminated by a carriage-return character, otherwise the program hangs!}
\vspace{12pt}

\section{How {\sc Combine} Works}\label{COMBINE_How Combine Works}

For a peak with given frequency and significance, all possible combinations of previously detected genuine frequencies $f_{k}$, $k=1,...,K$ are computed. $K$ is the maximum number of frequencies in a linear combination. The resulting frequency for a linear combination is
\begin{equation}
f^{\prime}:=\sum_{k=1}^{K}c_{k}f_{k}
\end{equation}
and shall be compared to a frequency $f$ in the input file.

\subsection{Sig vs.~csig}\label{COMBINE_Sig vs. csig}

If the keyword {\tt csig}\label{COMBINE_keyword.csig} is provided in the file {\tt <infile>.ini}, the cumulative sig (Reegen 2007, 2009) is used instead of the sig. This keyword does not require any parameters.

\subsection{Frequency resolution}\label{COMBINE_Frequency resolution}

The adjustment of the frequency resolution $\delta f$ is consistent with Eq.\,\ref{CINDERELLA_EQ fres} in the {\sc Cinderella} manual (Reegen 2009), where the total time interval width $T$ has to be provided by the user, because the time series is not incorporated by {\sc Combine}. Moreover, the user is more flexible if allowed to specify a different value for $T$. This interval width is provided by means of the keyword {\tt dt}\label{COMBINE_keyword.dt} in the file {\tt <infile>.ini}, followed by a floating-point number. The default setting is that {\sc Combine} determines the closest pair of frequencies and uses its inverse frequency spacing as $T$.

The second parameter, $\tau$, is specified using the keyword {\tt tol}\label{COMBINE_keyword.tol}, again followed by a floating-point number, in full consistency with {\sc Cinderella}. The default value is $\tau = 0$, forcing {\sc Combine} to employ the Rayleigh frequency resolution.

The frequency tolerance permits linear combinations where
\begin{equation}\label{COMBINE_EQ acc}
\alpha := \left| f-f^{\prime}\right|\le\delta f
\end{equation}
only. The quantity $\alpha$ is the accuracy of a linear combination and provided in the output.

\subsection{Limit of harmonic order}\label{COMBINE_Limit of harmonic order}

The range of harmonic orders is restricted by the parameter N, which is calculated according to
\begin{equation}\label{COMBINE_EQ N}
N=\mathrm{ceil}\left(\sqrt{\Omega\frac{\mathrm{sig}_{k}}{\mathrm{sig}_{K}}}\right)\: ,
\end{equation}
where $\mathrm{sig}_{k}$ denotes the sig associated to the frequency $f_{k}$ and $\mathrm{sig}_K$ is the sig associated to the last frequency in the input file, $f_K$. If the keyword {\tt csig} is set, the csig is consistently taken instead of the sig. The parameter $\Omega$ is provided by the keyword {\tt order}\label{COMBINE_keyword.order} in the file {\tt <infile>.ini}, followed by a floating-point number. The default value is $1$. Given the limit $N$, the coefficients of a linear combinations are restricted to indices from $-N$ to $N$ according to
\begin{eqnarray}
c_{k}=-N,\ldots,N
\end{eqnarray}.

\subsection{Equivalent sig}\label{COMBINE_Equivalent sig}

Each linear combination is assigned an equivalent sig,
\begin{equation}\label{COMBINE_EQ sigeq}
\mathrm{sig}_{\mathrm{eq}}:=\min\left(\left|c_{k}\right|^{-\delta_{k}}\mathrm{sig}_{k}\right)-\chi \log K\: ,
\end{equation}
where $\delta_{k}$ denotes the decay parameter provided by the keyword {\tt decay}\label{COMBINE_keyword.decay}, and $\chi$ is the combination damping, specified using the keyword {\tt cdamp}\label{COMBINE_keyword.cdamp}. Both keywords are followed by floating-point numbers. The default values for both parameters are $1$.

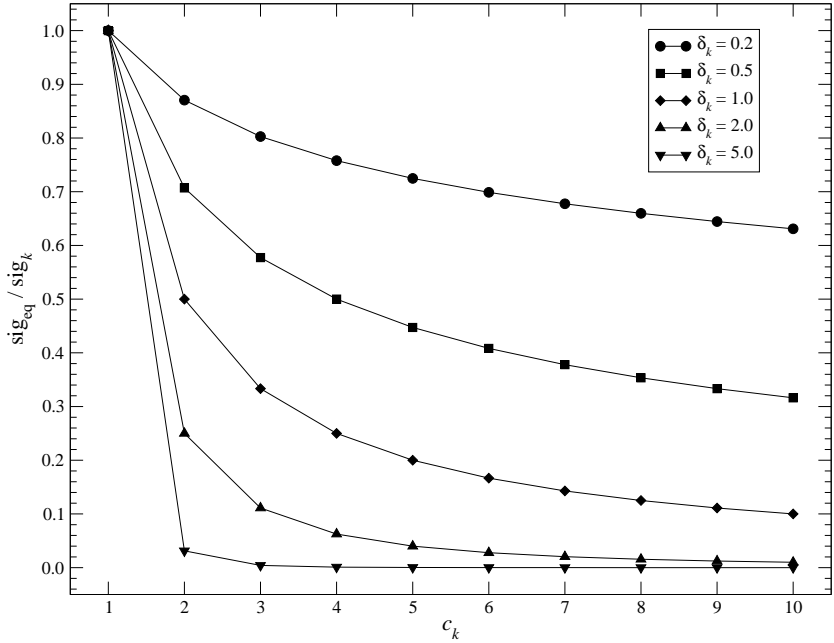
\figureDSSN{f1.eps}{Ratio of equivalent sig over sig of an individual signal component vs.~polynomial coefficient $c_k$ associated to the signal component. Five graphs for different values of the decay parameter $\delta _k$ are presented.}{COMBINE_decay}{!htb}{clip,angle=0,width=110mm}

Fig.\,\ref{COMBINE_decay} displays the relative sig correction with increasing coefficient $c_k$ for five different values of the decay parameter $\delta _k$. Fig.\,\ref{COMBINE_cdamp} illustrates the correction of equivalent sig with increasing number of components contributing to a linear combination $K$ for five different values of the combination damping $\chi$. 

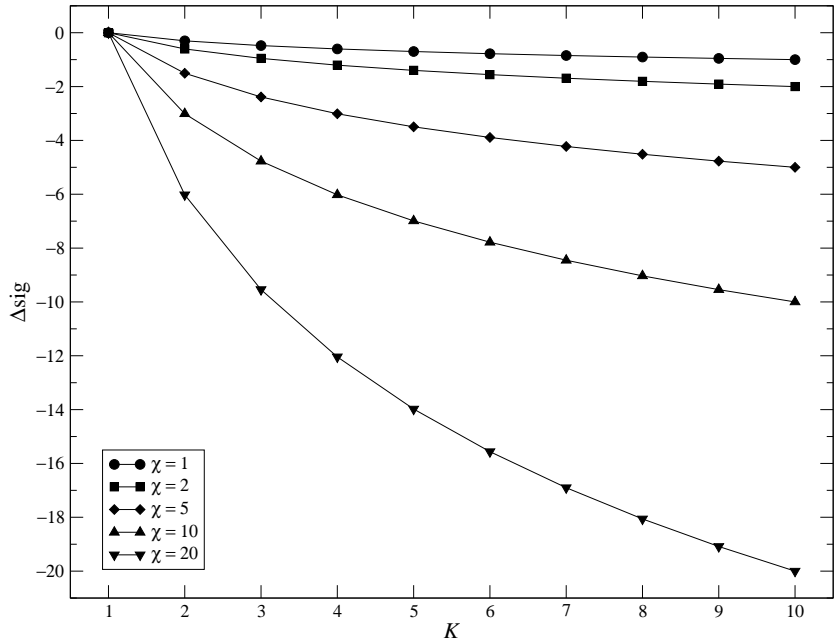
\figureDSSN{f2.eps}{Additive significance correction for a linear combination employing $K$ different signal components. Five graphs for different values of the combination damping $\chi$ are presented.}{COMBINE_cdamp}{!htb}{clip,angle=0,width=110mm}

\subsection{Reliability and sensitivity}\label{COMBINE_Reliability and sensitivity}

A linear combination is only accepted if the equivalent sig of the combination is high enough compared to the significance of the given peak according to
\begin{equation}\label{COMBINE_EQ relsens}
R := \frac{\mathrm{sig}_{\mathrm{eq}}}{\mathrm{sig}_{f}}\ge S\: ,
\end{equation}
where $S$ is the sensitivity, which can be adjusted by means of the keyword {\tt sens}\label{COMBINE_keyword.sens} in the file {\tt <infile>.ini}. The keyword is followed by a floating-point number, and the default value is $0.1$. If all examined linear combinations have a reliability below $S$, the examined signal component is considered genuine. Hence the sensitivity provided by the keyword {\tt sens} permits to directly adjust the number of genuine components in a list of frequencies.

The ratio of sigs, $R$, is called the reliability of a linear combination and part of the {\sc Combine} output. If multiple combinations are available, the reliability is used to decide which one to pick. This means, {\sc Combine} picks the combination with the highest reliability.

\section{Output}\label{COMBINE_Output}

Genuine frequencies are assigned identifiers {\tt f\#index\#}, where {\tt \#index\#} denotes an integer number starting at $1$. According to the number of significant signals present in the file {\tt <infile>}, {\sc Combine} chooses a constant number of digits. For example, if the input file contains from $1$ to $9$ frequencies, the identifiers for genuine frequencies are {\tt f1}, {\tt f2}, ... If the input file contains from $10$ to $99$ frequencies, {\sc Combine} enumerates the genuine components {\tt f01}, {\tt f02}, ..., and so on. This format convention applies to the indexing of rows also.

Linear combinations are denoted by the frequency identifiers of the genuine components and appear as a formula: if the frequency under consideration is, e.\,g., $f_1+3f_3-2f_{10}-f_{14}-0.00214$, {\sc Combine} displays it as {\tt f01+3f02-2f10-f14-0.00214} both on the screen and in the output file. In this context, $-0.00214$ is the frequency accuracy.

The screen output consists of a single line for each signal (i.\,e., for each row in the input file). {\sc Combine} displays
\begin{enumerate}
\item the row index,
\item the linear combination including the frequency accuracy, and
\item the reliability $R$ (Eq.\,\ref{COMBINE_EQ relsens}).
\end{enumerate}
For genuine frequencies, {\sc Combine} displays only the row index and the frequency identifier. At runtime, the most reliable linear combination identified so far is displayed. If {\sc Combine} finds a ``better'' solution, the line on the screen is updated.

By default, {\sc Combine} generates an output file {\tt <infile>.cmb}. It contains a row index in the first column, then all information of the input file in the further columns, plus three additional columns at the end:
\begin{enumerate}
\item reliability $R$ (Eq.\,\ref{COMBINE_EQ relsens})\footnote{Zero values indicate genuine frequencies},
\item total number of linear combinations within the frequency resolution,
\item the linear combination itself, plus the frequency accuracy. If a frequency is considered genuine, only the frequency identifier is displayed.
\end{enumerate}

For convenience, a second output file {\tt <infile>.gen} is produced by {\sc Combine}. It is truncated to the genuine frequencies only and contains the row index in the first column, then all the information provided in the input file, plus the frequency identifier in the last column. The columns for the reliability and the number of linear combinations within the frequency resolution are omitted. This file provides the opportunity to have all the genuine frequencies available at a glance.

\vspace{12pt}\noindent{\bf Example.}\footnote{The computation of the sample project {\tt CombineNative} takes 40 minutes on an Intel Core2 CPU T5500 (1.66GHz) under Linux 2.6.18.8-0.9-default i686.} \it The sample project {\tt CombineNative} contains a list of significant frequencies found in the MOST\footnote{MOST is a Canadian Space Agency mission, jointly operated by Dynacon Inc., the University of Toronto Institute of Aerospace Studies, the University of British Columbia, and with the assistance of the University of Vienna, Austria.} (Microvariability \& Oscillations of STars) photometry of $\zeta$ Oph (Walker et al.~2003, 2004, 2005). According to the input file {\tt result.dat}, altogether 294 formally significant signal components (sig $>$ 5) were identified.

The file {\tt result.dat.ini} contains five keywords:

\begin{scriptsize}\begin{verbatim}
order 0.2
dt 26
decay 1.5
cdamp 10
sens 0.2
\end{verbatim}\end{scriptsize}

\noindent The dataset is 26 days long, and the frequencies are provided in cycles per day. Thus {\sc Combine} will assume a Rayleigh frequency resolution of 0.03846 cycles per day. There is no specification for the frequency tolerance parameter (keyword {\tt tol}). Thus the default setting 0 is used.

Running {\sc Combine} by typing the command line {\tt Combine result.dat} yields a welcome message on the screen.

\begin{scriptsize}\begin{verbatim}
 CCCCCC                   bb      ii                 
CC    CC                  bb                         
CC        ooooo  m mm mm  bb bbb  ii n nnnn   eeeee  
CC       oo   oo mm mm mm bbb  bb ii nn   nn ee   ee 
CC       oo   oo mm mm mm bb   bb ii nn   nn ee   ee 
CC       oo   oo mm mm mm bb   bb ii nn   nn eeeeeee 
CC       oo   oo mm mm mm bb   bb ii nn   nn ee      
CC    CC oo   oo mm mm mm bb   bb ii nn   nn ee   ee 
 CCCCCC   ooooo  mm mm mm b bbbb  ii nn   nn  eeeee  


Version 1.0
************************************************************
by Piet Reegen
Institute of Astronomy
University of Vienna
Tuerkenschanzstrasse 17
1180 Vienna, Austria
Release date: August 18, 2009
\end{verbatim}\end{scriptsize}

The program finds out that the input file is a seven-column {\sc SigSpec} result file, determines the number of rows and reads the input data. Note that 295 rows correspond to 294 significant signal components, because the last row in the {\sc SigSpec} result file contains information on the residuals (see {\sc SigSpec} manual, p.\,\pageref{SIGSPEC_Result files}).

\begin{scriptsize}\begin{verbatim}
*** start **************************************************

File result.dat: SigSpec format
rows                                    295
read input file
\end{verbatim}\end{scriptsize}

Then the search for linear combinations starts. For each row in the input file, {\sc Combine} displays the most reliable combination detected so far.

The first four signal components are found to be genuine. Since the number of signal components is 294, {\sc Combine} uses a three-digit format for the row indices and frequency identifiers.

\begin{scriptsize}\begin{verbatim}
row 001: f001
row 002: f002
row 003: f003
row 004: f004
\end{verbatim}\end{scriptsize}

For rows 5 and 6 in the input data, the screen output contains the most reliable linear combination (including the frequency accuracy) and the reliability.

\begin{scriptsize}\begin{verbatim}
row 005: 3f001-f002-2f003-f004+0.0284306 0.236585
row 006: 3f001+2f002-f004+0.0136421 0.35803
\end{verbatim}\end{scriptsize}

An examination of the output file {\tt result.dat.cmb} shows that rows {\tt 005} and {\tt 006} end with

\begin{scriptsize}\begin{verbatim}
0.2365853347754522  1 3f001-f002-2f003-f004+0.0284306168856169
0.3580304203945811  2 3f001+2f002-f004+0.0136420746028509
\end{verbatim}\end{scriptsize}

\noindent These entries refer to the columns added by {\sc Combine}. The first value is the reliability, the second one is the number of examined linear combinations, and the last column represents the linear combination itself. For row {\tt 005}, there is only one linear combination available within the frequency resolution, for row {\tt 006} the number of linear combinations taken into account is 2.

Subsequently, the screen output indicates a fifth genuine frequency.

\begin{scriptsize}\begin{verbatim}
row 007: f005
\end{verbatim}\end{scriptsize}

The frequency in row number 8 is 0.02783 cycles per day, which is below the frequency resolution. Thus the component is considered to refer to zero frequency, and in this case, no reliability is evaluated.

\begin{scriptsize}\begin{verbatim}
row 008: 0+0.0278395
\end{verbatim}\end{scriptsize}

In the further rows of the input files, no more genuine frequencies are detected.

\begin{scriptsize}\begin{verbatim}
row 009: -f002+f005-0.025485 0.759005
row 010: f001-f002-f004+f005+0.0313392 0.490535
row 011: -f001+f004-0.00275538 1.26888
row 012: f001-f002-f004+f005-0.0295542 0.680494
row 013: -2f001+2f003+f004-0.00567519 0.523911
row 014: -f001+f005+0.024731 1.72772
row 015: 2f002+0.0249392 1.47442
row 016: 2f001-f004-0.0100088 1.70761
row 017: -f001+2f002-0.00217389 1.55951
row 018: f001-f002+0.00824894 3.95466
row 019: f002+f005-0.00668728 1.64167
row 020: 2f002+f003-f005-0.00199182 0.779607
\end{verbatim}\end{scriptsize}

It is a remarkable matter of fact that {\sc Combine} is able to compose all 294 frequencies contained by the input file as linear combinations of no more than five genuine frequencies. However, a different parameter constellation in the configuration file {\tt result.dat.ini} can produce completely different output. Note that the time consumption by {\sc Combine} dramatically increases with the number of genuine frequencies identified. This is because more genuine frequencies increase the number of possible linear combinations over-proportionally. A list of genuine frequencies only is found in the output file {\tt result.dat.gen}.\sf

\begin{scriptsize}\begin{verbatim}
5 genuine frequencies found.


Finished.

************************************************************

Thank you for using Combine!
Questions or comments?
Please contact Piet Reegen (reegen@astro.univie.ac.at)
Bye!
\end{verbatim}\end{scriptsize}

\section{Order of Input Rows}\label{COMBINE_Order of Input Rows}

Since {\sc Combine} processes the input file row by row, the order of rows plays a crucial part in the way the analysis is performed. Changing the order of rows in the input file influences the base upon which the linear combinations are formed. Thus, if there are frequencies previously known to be genuine, it is advisable to ensure that they are on top of the input file, if all further frequencies are supposed to be checked for linear combinations of preferrably these components.

\vspace{12pt}\noindent{\bf Example.}\footnote{The computation of the sample project {\tt order} takes 40 minutes on an Intel Core2 CPU T5500 (1.66GHz) under Linux 2.6.18.8-0.9-default i686.} \it The input of the sample project {\tt order} is essentially the same as for {\tt CombineNative}. Only the order of rows is slightly modified: the 6th signal component of the file {\tt result.dat} in the project {\tt CombineNative}, which refers to the orbit frequency of the MOST spacecraft, appears now on top. This re-ordering forces {\sc Combine} to consider $14.188\,\mathrm{d}^{-1}$ genuine. Also the configuration file {\tt result.dat.ini} is the same as for the project {\tt CombineNative}.

Again, there is a base of five genuine frequencies three of which are identical to the project {\tt CombineNative}, namely 5.182, 2.675 and 3.055 cycles per day. The two genuine signal components at 6.722 and 7.193 cycles per day are replaced by 14.188 and 0.0697 cycles per day.\sf

\section{Rejecting Unwanted Linear Combinations}\label{COMBINE_Rejecting Unwanted Linear Combinations}

Moreover, the user may indicate unwanted signal components in the input file {\tt <infile>} by applying a minus sign to the corresponding frequencies. {\sc Combine} reacts with a corresponding change of the sign for the reliability. If the user additionally provides the keyword {\tt reject}\label{COMBINE_keyword.reject} in the file {\tt <infile>.ini}, all rows are rejected from the output file {\tt <infile>.cmb} for which the most reliable linear combination contains one or more unwanted frequencies.

The screen output contains linear combinations incorporating unwanted frequencies at runtime. To indicate such unwanted combinations, the reliability is displayed as a negative value. If the examination of an input line finishes with the ``best'' linear combination containing an unwanted frequency, the corresponding line is removed from the screen output.

\vspace{12pt}\noindent{\bf Example.}\footnote{The computation of the sample project {\tt reject} takes 40 minutes on an Intel Core2 CPU T5500 (1.66GHz) under Linux 2.6.18.8-0.9-default i686.} \it The input of the sample project {\tt reject} is the same as for {\tt order}, with a minus sign for the first frequency of 14.188 cycles per day, which represents the orbit of the MOST spacecraft. The file {\tt result.dat.ini} contains an additional line,

\begin{scriptsize}\begin{verbatim}
reject
\end{verbatim}\end{scriptsize}

\noindent The combination of this keyword and the negative sign for the first signal component in the input file forces {\sc Combine} to reject all linear combinations incorporating the frequency 14.188 cycles per day from the output file {\tt result.dat.cmb}. In the screen output, such linear combinations are indicated by a negative reliability, e.\,g.

\begin{scriptsize}\begin{verbatim}
row 005: f001+3f002+2f003+0.0136421 -0.325575
\end{verbatim}\end{scriptsize}

This entry is visible at runtime, but vanishes from the screen output when the calculations for row 006 start.
\sf

\section{Keywords Reference}\label{COMBINE_Keywords Reference}

This section is a compilation of all keywords accepted by {\sc Combine}. A brief description of arguments and default values is given. If an argument is required, it is indicated by {\tt <double>}, and default values are given in parentheses, e.\,g.~{\tt (1)}.

\subsubsection{\tt cdamp <double> (1)}

combination damping, e.\,g. reduction of reliability of a linear combination with increasing number of components employed, p.\,\pageref{COMBINE_keyword.cdamp}

\subsubsection{\tt csig}

forces {\sc Combine} to use csig instead of sig, p.\,\pageref{COMBINE_keyword.csig}

\subsubsection{\tt decay <double> (1)}

decay of reliability assigned to a frequency multiple for increasing harmonic order, p.\,\pageref{COMBINE_keyword.decay}

\subsubsection{\tt dt <double> (auto)}

total time interval of the time series, defining the Rayleigh frequency resolution. By default, {\sc Combine} determines the Rayleigh frequency resolution as the frequency spacing of the closest pair of frequencies found in the input data, p.\,\pageref{COMBINE_keyword.dt}.

\subsubsection{\tt order <double> (auto)}

parameter restricting the range of harmonics of individual frequency components to be employed to form linear combinations, p.\,\pageref{COMBINE_keyword.order}

\subsubsection{\tt reject}

activates the rejection of unwanted linear combinations. The user may indicate unwanted frequencies by a minus sign in the input file {\tt <infile>}. If this keyword is set, {\sc Combine} automatically suppresses the output of those signal components for which the most reliable linear combination incorporates such an unwanted frequency, p.\,\pageref{COMBINE_keyword.reject}.

\subsubsection{\tt sens (0.1)}

reliability limit to be exceeded in order to accept a linear combination, adjusts the number of genuine components in a frequency list, p.\,\pageref{COMBINE_keyword.sens}

\subsubsection{\tt tol <double> (0)}

{\sc Combine} frequency tolerance parameter, p.\,\pageref{COMBINE_keyword.tol}

\section{Online availability}

The ANSI-C code is available online at {\tt http://www.sigspec.org}. For further information, please contact P.~Reegen, {\tt peter.reegen@univie.ac.at}.

\pagebreak

\acknowledgments{PR received financial support from the Fonds zur F\"or\-de\-rung der wis\-sen\-schaft\-li\-chen Forschung (FWF, projects P14546-PHY, P17580-N2) and the BM:BWK (project COROT). Furthermore, it is a pleasure to thank
M.~Gruberbauer, M.~Hareter, D.~Huber, D.\,Punz (Univ.~Vienna), G.\,A.\,H.~Walker (UBC, Vancouver), and W.\,W.~Weiss (Univ.~Vienna) for their help. I acknowledge the anonymous referee for a detailed examination of both this publication and the corresponding software, as well as for the constructive feedback that helped to improve the overall quality a lot. Finally, I address my very special thanks to J.\,D.~Scargle for his valuable support.
}

\References{

Kallinger, T., Reegen, P., Weiss, W.\,W.~2008, A\&A, 481,  571\\

Reegen, P.~2005, in {\it The A-Star Puzzle}, Proceedings of IAUS 224, eds. J. Zverko, J. Ziznovsky, S.J. Adelman, W.W. Weiss (Cambridge: Cambridge Univ.~Press), p.~791\\

Reegen, P.~2007, A\&A, 467, 1353\\

Reegen, P.~2009, CoAst, submitted\\

Reegen, P., Gruberbauer, M., Schneider, L., Weiss, W.\,W.~2008, A\&A, 484, 601\\

Walker G., Matthews J., Kuschnig R., et al.~2003, PASP, 115, 1023

Walker G.\,A.\,H., Matthews, J.\,M., Kuschnig, R., et al.~2004, BAAS, 36, 1361

Walker G.\,A.\,H., Kuschnig, R., Matthews, J.\,M., et al.~2005, ApJ, 623, L145

}


\setlength{\columnseprule}{0pt}\clearpage
\thispagestyle{empty}
\cleardoublepage
\thispagestyle{empty}

\end{document}